# Why early tactile speech aids may have failed: no perceptual integration of tactile and auditory signals

Aurora Rizza[1], Alexander V. Terekhov[2], Guglielmo Montone[2], Marta Olivetti Belardinelli[1,2,3], and  J. Kevin O'Regan[2]

[1]Laboratoire Psychologie de la Perception,  Université Paris Descartes, France
[2] Psychological Department, Faculty of Medicine and Psychology, Sapienza, University of Rome, Italy
[3]*ECONA Interuniversity Centre for the Research on Cognitive Processing in Natural and Artificial Systems, Rome, Italy*

**Correspondence:**
Aurora Rizza
aurorarizza@gmail.com

**Abstract**
Tactile speech aids, though extensively studied in the 1980's and 90's, never became a commercial success. A hypothesis to explain this failure might be that it is difficult to obtain true perceptual integration of a tactile signal with information from auditory speech: exploitation of tactile cues from a tactile aid might require cognitive effort and so prevent speech understanding at the high rates typical of everyday speech. To test this hypothesis, we attempted to create true perceptual integration of tactile with auditory information in what might be considered the simplest situation encountered by a hearing-impaired listener. We created an auditory continuum between the syllables /BA/ and /VA/, and trained participants to associate /BA/ to one tactile stimulus /VA/ to another tactile stimulus. After training, we tested if auditory discrimination along the continuum between the two syllables could be biased by incongruent tactile stimulation. We found that such a bias occurred only when the tactile stimulus was above, but not when it was below its previously measured tactile discrimination threshold. Such a pattern is compatible with the idea that the effect is due to a cognitive or decisional strategy, rather than to truly perceptual integration. We therefore ran a further study (**Exp 2**), where we created a tactile version of the McGurk effect. We extensively trained two Subjects over six days to associate four recorded auditory syllables with four corresponding apparent motion tactile patterns. In a subsequent test, we presented stimulation that was either congruent or incongruent with the learnt association, and asked Subjects to report the syllable they perceived. We found no analog to the McGurk effect, suggesting that the tactile stimulation was not being perceptually integrated with the auditory syllable. These findings strengthen our hypothesis according to which tactile aids failed because integration of tactile cues with auditory speech occurred at a cognitive or decisional level, rather than truly at a perceptual level.

## 1    Introduction

Efforts in 1970-1990's to create tactile speech aids for hearing impaired listeners were promising and gave improvements in speech perception similar to the new technology of cochlear implants being developed at that time (Osberger et al. 1991; Carney et al., 1993; Sarant et al. 1996). However subsequent research on tactile aids did not live up to expectation. Interest in tactile speech aids subsided and cochlear implants became the dominant technology we know today. Yet there are cases when cochlear implants may be too expensive or medically counterindicated, and a more affordable and less invasive tactile auditory supplementation device would present a viable alternative. It is therefore important to understand what the problem was with the tactile speech aids.

It might be thought that reasons for the failure of tactile aids resided in the great effort needed to learn to use them, in the practical problems involved in reliably providing useful information through the interface of the skin, and in the fact that the devices were cumbersome. However analogous problems also burdened cochlear implants. Worse, cochlear implants involved the considerable onus of expensive surgical procedures. Yet for cochlear implants the problems were overcome.

In this article we wish to investigate a deeper explanation for the lack of success of tactile hearing aids. It is the idea that early tactile speech aids never provided proper perceptual tactile-to-audio integration. Proper perceptual integration is needed in an effective tactile aid so that the tactile information provided by the device is combined with the auditory signal in an automatic way. Only if integration is automatic and effortless, can it complement or even replace the audio speech signal when speech is delivered at the fast rate that characterizes normal social interactions.

The idea that proper perceptual tactile-to-audio integration might not have been obtained in tactile aids at first seems surprising given the large amount of contemporary literature on cross-modal interactions suggesting the existence of auditory-tactile integration for non-linguistic stimuli, and some research suggesting the same for speech.

Indeed, a cursory examination of the literature would suggest that crossmodal interactions between audio and tactile modalities are widespread, at both behavioral (e.g. Yau et al., 2009, 2010; Jousmäki & Hari, 1998; Ro et al., 2009, Foxe et al., 2009; Olivetti Belardinelli, 2011) and neural levels (e.g. Kassuba et al., 2013; Caetano and Jousmäki, 2006; Foxe et al., 2002). Specific evidence for an influence going from tactile to auditory perception (rather than the reverse) would seem to come, for example, from Gillmeister & Eimer (2007), who showed that irrelevant tactile stimulation affects the perception of the loudness of a sound; and from Yau et al. (2010), who showed that tactile distractors can influence judgments of auditory intensity. Similarly, Schürmann et al. (2004), Yarrow et al. (2008) and Okazaki et al. (2012) showed that the perceived loudness of an auditory tone seemed to increase when subjects were holding a vibrating object, also suggesting a facilitation of the auditory signals by tactile stimulation. Soto-Faraco et al. (2004) also showed that apparent tactile motion can influence auditory motion judgments. Finally, in the context of cochlear implants, Nava, Bottari, Villwock, Fengler, Büchner, Lenarz, & Röder (2014) showed that the presence of an auditory pulse could be better detected when accompanied by a tactile pulse.

As concerns knowing whether tactile inputs might complement *speech* perception, apparently supportive claims come from Gick & Derrick (2009), who showed that syllables heard simultaneously with cutaneous air puffs were more likely to be heard as aspirated (for example, causing participants to mishear 'b' as 'p'). Alcantara et. al (1993), Cowan et al. (1990), and Galvin et al. (2001) reported an advantage of using both auditory and tactile modalities during training with a vibrotactile speech aid, as compared to using unimodal auditory, tactile or visual training. Finally, other research has also shown tactile facilitation of auditory information in speech using the Tadoma method (Alcorn, 1932), and in music (Darrow, 1989; Calabrese & Olivetti Belardinelli, 1997).

However a problem exists in the interpretation of the majority of the studies cited above: In these studies it is not clear whether the observed effects of facilitation are due to truly perceptual integration. For example, in Cowan et al.'s (1990) study, the authors trained fourteen prelingually profoundly hearing-impaired children using a training program that combined tactile and tactile/auditory feature recognition exercises with conversational combined-modality (tactual, auditory, and visual) games designed to encourage integration of learned feature-level information into conversational tasks. Results for three speech tests showed significant improvement when the tactile aid was used in combination with hearing aids as compared with hearing aids alone and as compared with lipreading alone, and hearing aids as compared with combined lipreading and hearing aids. This result shows that additional information was provided by the tactile signal over and above what was provided by the auditory signal, but we have no proof that the integration was truly perceptual. Instead, it might have occurred at what might be called the decisional, cognitive or response-level.

Another case in point is Jousmaki & Hari's (1998) well-established parchment-skin illusion, in which an observer's sensed moistness/roughness of their skin while rubbing their hands together is influenced by concomitant auditory stimulation. As noted by Soto-Faraco and Deco (2009) and Gescheider et al (1970, 1974), it is difficult to separate the perceptual and cognitive contributions to this effect. This is because the cues used to influence perceptual judgments were provided above the perceptual threshold. It can thus be argued that the experienced percept itself was not changed by the additional cue, and that instead, the observer simply used the knowledge obtained from the perceptually available cue to modify their interpretation or response.

From these examples and similar criticisms that can be levelled at other studies, we see that the relative contribution of perceptual versus cognitive (or decisional) processes to multisensory integration remains a problem in the interpretation of existing results on tactile-auditory integration. The hypothesis thus becomes more plausible that the reason that previous tactile speech prostheses were not adopted is that the tactile information provided to listeners could not be perceptually integrated with the auditory speech signal. This would mean that substantial cognitive effort would be necessary to make use of the tactile information in disambiguating the audio. Speech processing would become difficult at normal speech rates, and the speech aid would be of limited use.

The purpose of our work was therefore to check whether proper perceptual integration of tactile information with a speech signal can actually be obtained. To maximise the chances of success, a first experiment used a very simple task, namely the task of distinguishing two phonemes when disambiguating tactile information is provided. We approximated the situation of a hearing impaired listener by low pass filtering the audio signal. We provided an amount of training similar to what was used in the existing literature on tactile aids. In a second experiment we provided much more extensive training and used a paradigm similar to the audio-visual McGurk effect.

## 2      Experiment 1: Psychophysical experiment

By splicing together parts of the syllables /BA/ and /VA/, we created an auditory continuum between the two syllables, and asked people to judge whether they heard /BA/ or /VA/. At the same time they were given tactile cues which we hoped they would integrate into their perceptual judgment. To verify that the effect was perceptual rather than cognitive, we adopted the reasoning of Jain et al. (2010), who points out that if a perceptual judgment can be influenced by a "weak" below-threshold cross-modal cue, then the influence of the cue is unlikely to be cognitive or decisional, and more likely to be truly perceptual. Thus in the test phase of the experiment we provided either no tactile cue, a "weak" below threshold tactile cue, or a "strong" above threshold tactile cue.

### 2.1     Methods

#### 2.1.1    Subjects
We tested 10 normal hearing subjects (8 males, 2 females, mean age 33.2). Subjects were undergraduates, Ph.D. students and postdoc researchers. Subjects belonged to different cultures and had different mother tongues.

#### 2.1.2    Apparatus
The tactile stimulation device consisted in two side-by-side dynamic braille stimulators (B11, Metec AG, Stuttgart, Germany) each with 2 x 4 pins, creating a 4x4 array. The device was held in the hand with the subject's thumb lying on the array (see Figure 1). The device was interfaced to a HP Elitebook 14" computer and controlled by a python program. The auditory stimuli were delivered through 'Marshall Monitor' headphones at a comfortable volume of approximately 60 dB for all subjects.

### 2.1.3 Stimuli

To create the auditory continuum going from the sound /VA/ to the sound /BA/ we first recorded these two syllables from a natural male voice. We then created 21 composite stimuli between /VA/ and /BA/ by attaching a fraction p from the beginning of the recording of /V/ with a fraction (1-p) from the end of the recording /B/ and then attaching this to the sound /A/ from /VA/[1]. The fraction p increased from 0 to 1 in equal steps. Other ways of morphing between /VA/ and /BA/ were also attempted using existing software[2], but proved to sound less natural than this method. We then applied a lowpass filter with a cutoff at 1850 Hz to each exemplar in order to degrade the auditory quality so that it approximated that experienced by a typical age-related hearing impaired listener (e.g. Huang & Tang, 2010), and yet allowed the stimuli on the endpoints of the continuum to be heard clearly with a gradual transition between the two endpoints. We labelled the files from "0" (pure /VA/) to "20" (pure /BA/). In a pilot study we found that people tended to have their threshold at around stimulus 14, and thus we centered the stimuli range we used in the experiment around it: from "8" to "20". The sound files can be found on
https://github.com/montoneguglielmo/stimuliFeelSpeechFrontier

For the tactile stimulation, we constructed 13 stimuli, in which the top line of pins stroked the thumb by moving at different speeds either from left to right or from right to left. The stimuli were constructed in the following way. Each pin stayed up for 20ms, and then it went down. To create the impression of stroking, we varied the time between the moment when successive pins went up. For example for a time of 0 ms, the pins went up all together and there was no apparent stroking motion. For a time of +40 ms each successive pin came up 40 ms after the beginning of the moment when previous pin came up, giving an apparent motion from left to right. For a time of -40 ms, the motion went in the opposite direction. We used times of -40ms, -30ms, -20ms, -15ms, -10ms, -5ms, 0ms, 5ms, 10ms, 15ms, 20ms, 30ms, 40ms. The movement of the pins makes a weak sound that cannot be heard when wearing the headphones.

### 2.1.4 Procedure

Each subject ran six parts: measurement of the auditory psychometric function, measurement of the tactile psychometric function, 9 blocks of auditory-tactile training, auditory-tactile test,

---

[1] Using the /A/ from the sound /BA/ or a mixture of the two proved less natural than just using the /A/ from /VA/.
[2] We tried STRAIGHT, developed by Prof. Hideki Kawahara's group at Wakayama University, and PRAAT, developed by P. Boersma and D. Weenink from University of Amsterdam

auditory-tactile test with below threshold tactile stimulation ("weak" bias condition), auditory-tactile test with above threshold tactile stimulation ("strong" bias condition), measurement of the auditory psychometric function "post-test".

*Auditory.* The measurement of the auditory psychometric function included a familiarization block and a main block. In the familiarization block, the subjects heard three stimuli (clear /VA/, clear /BA/ and the middle exemplar half way between the extremes), each one repeated three times (in random order) and they had to answer whether they heard /BA/ or /VA/. Participants responded by pressing the buttons '1' or '2' of the keyboard with the non-dominant hand. Subjects pressed the key '1' when they heard /BA/ and the key '2' when they heard /VA/. In the main block, we presented all 13 auditory stimuli in random order (each of them was presented 15 times), and subjects responded using the same keys to indicate whether they heard /BA/or /VA/. In this phase, we obtained a psychometric function for each subject, and determined its mean [μ] (the stimulus that the subject perceived 50% of the time as /BA/ or /VA/), that is, the subject's point of subjective equality (PSE) and the measure of the slope or deviation [σ] (see Figure 2 for examples from two subjects, and Supplementary Material for all subjects).

*Tactile.* The measurement of the tactile psychometric function was similar to that of the auditory. It included two familiarization blocks and a main block. We asked the subject to press the key '1' when the subject felt that the pins moved to the left and to press the key '2' when the subject felt that the pins moved to the right. In the first familiarization block we presented the pins moving to the left with different speeds (-40ms, -30ms, -20ms, .-15ms, -10ms, -5ms), each speed repeated twice in random order, while in the second familiarization block, we presented the pins moving to the right with different speeds (5ms, 10ms, 15ms, 20ms, 30ms, 40ms) each speed repeated twice in random order. In the main block we presented the 12 stimuli from the familiarization block, plus a stimulus with 0ms delay, in random order (each stimulus was presented 15 times), and subjects indicated in the same way as before whether they felt the motion going from left to right or opposite. We computed psychometric functions for each subject, and determined their means [μ], corresponding to the subject's point of subjective equality (PSE), and slopes [σ] (see Figure 2 for examples from two subjects, and Supplementary Material for all subjects).

*Audio-tactile training.* After the measurement of the auditory-only and tactile-only psychometric functions, the subjects performed the audio-tactile part of the experiment. During this part the auditory syllables were presented simultaneously with tactile stimuli. The syllables on the "/BA/ side" of the subject's auditory PSE were presented together with the tactile stimulus "left", and the stimuli on the "/VA/ side" were presented together with the tactile stimulus "right". The tactile stimuli "left moving" or "right moving" were generated by setting the delay to the value PSE - xσ and PSE + xσ, where x varied between 0.5 and 3 and determined how clearly perceivable the tactile stimulus was to be.

In the familiarization block, we used the two extreme /BA/ and /VA/ auditory stimuli that we used for the familiarization in the measurement of the psychometric function in the auditory-only condition, each one repeated three times. The extreme cases of /BA/ and /VA/ were each associated with two clearly distinguishable tactile stimuli, corresponding to the tactile stimuli that were at ±3σ from the PSE tactile stimulus.

For the 9 main blocks of auditory-tactile training, Subjects were presented with all the 13 auditory stimuli; the "left moving" tactile stimuli were associated with the auditory stimuli that were on the "/BA/ side" of the auditory PSE measured for each subject, and the "right moving" tactile stimuli were associated with the auditory stimuli that were on the "/VA/ side" of the auditory PSE for that subject. However we changed the precise parameters of the two tactile stimuli that were associated with these two sets of auditory stimuli across the blocks -- first and second block: ±3σ of the tactile PSE, third block: ±2σ of the tactile PSE, fourth block: ±1.5σ of the tactile PSE, fifth, sixth, seventh blocks: ±1σ of the tactile PSE, eighth and ninth blocks: ±0.5σ of the tactile PSE (see Figure 3 for examples). Subjects had a break after each block.

*Audio-tactile test.* The "test" auditory-tactile block was identical to the last session of the audio-tactile training, i.e. the tactile stimulation was at ±0.5 σ of the tactile PSE. In this test block subjects were told that tactile stimulation could be misleading and that they should ignore it.

In the "weak bias" auditory-tactile test, we introduced a bias in the association between auditory and tactile stimuli. The auditory stimuli above PSE + 3 (three points above the PSE in the direction of /BA/) were mapped into tactile stimulation "left" and the stimuli below this value were mapped into tactile stimulation "right" (see Figure 4). The tactile stimulation used corresponded to the stimulations at ±0.5 σ from the tactile PSE. Note again that we asked the subjects to ignore the tactile inputs.

In the "strong bias" auditory-tactile test we linked auditory stimuli to tactile stimuli with the same bias as in the biased with weak tactile stimulation. However the tactile stimulation used was at ±2.0 σ from the tactile PSE (see Figure 4). Note again that we asked the subjects to ignore the tactile inputs.

The particular tactile stimuli that were used for the weak and strong bias conditions were selected to be respectively clearly below and clearly above the tactile threshold. In agreement with common practice for two-alternative forced choice task the threshold is assumed to be the stimulus for which the subject gives correct response in 75% of trials. A tactile stimulation of PSE - 2.0 σ or PSE + 2.0 σ will be correctly reported as left or right in about 98% of cases, which is thus clearly above threshold. Similarly, the tactile stimuli at PSE + 0.5σ or PSE - 0.5σ will be reported correctly in about 69% of cases, which is clearly below the threshold.

In a last, "auditory post-test" phase of the experiment we measured the auditory psychometric function again without tactile stimulation as a control condition to check that any observed change in audio-tactile PSE was really due to the tactile influence rather than some kind of auditory adaptation.

## 2.2    Data analysis

We measured the number of times a participant responded /BA/ or /VA/ to each stimulation in each block. The frequencies of responses were fitted with cumulative normal distributions using the PsychoPy toolbox, and their mean μ and slope σ determined. (see Figure 2 for examples from two subjects, and Supplementary Material for all subjects).

A repeated measures ANOVA was used to assess the effect of the condition on the measured PSE in the audio and audio-tactile conditions. The factors used in the ANOVA were: "pre-test", "test", "weak", "strong", "post-test". Paired t-tests were used to test the differences between individual conditions. The analysis was performed in R using the "aov" and "t.test" functions.

## 2.3    Results

### 2.3.1    PSE's

Figure 5 shows a plot of the /BA/-/VA/ PSEs for each subject in the pre-, post- and experimental conditions, and the means of these PSEs over all subjects. We were expecting that the PSEs for the biased conditions with 'weak' and 'strong' tactile stimulation would both be significantly higher than in the non-biased conditions (i.e. the auditory pre- and post-tests, and the test condition with no bias). However this was not evident from the statistics. Though the ANOVA revealed a significant effect of conditions on the measured PSEs ($F(4,36)=3.57$; $p<0.05$) the planned a posteriori comparisons did not provide clear results. We did find a statistically significant difference between the strong bias condition and the auditory pre- ($t(9)=-4.49$,; $p<0.005$) and post-tests ($t(9)=-2.38$, $p<0.05$), with the tactile bias causing a displacement of the auditory PSE in the direction of the bias. However, curiously, the difference was not significant as compared to the no-bias test condition ($t(9)=-1.67$, $p=0.13$). As concerns the weak bias condition, this was not significantly different from the other conditions.

Given these results, the tactile stimulation seems not to strongly influence the auditory PSE, except possibly in the "strong" condition, when it is presented clearly above threshold (-2.0 σ and +2.0 σ from the tactile PSE).

### 2.3.2 Subjective reports

Although subjects during the experiment were given only a forced choice between the responses 'BA' or 'VA', we asked subjects after the experiment to give any subjective impressions they had concerning the stimuli. We were expecting that in the training phase and the unbiased tests, subjects would hear the stimuli more clearly thanks to the tactile stimulation. However this was not evident from the subjective reports. On the other hand after several blocks of training, or during the test conditions, some subjects said they heard sounds intermediate between /BA/ and /VA/, like for example /VGA/ or /NGA/. Additionally almost all subjects reported hearing sounds that were completely different from the stimulus sounds. For example, one subject said he often heard the sound "clack", and another said he heard the sound "eco". This kind of effect occurred also in the exclusively auditory post-test, and so we hypothesized that it could be due to selective speech adaptation (Eimas and Corbit, 1973), a purely auditory phenomenon known to occur for repeated presentation of syllables. We further confirmed in an independent pilot test that a subject who ran the experiment with no tactile stimulation at all also experienced such auditory deformations.

## 2.4 Discussion

This experiment was designed to test if it is possible to obtain truly perceptual facilitation of auditory speech information by the use of a learned auditory-tactile association. Unfortunately our results are not clearcut. First, the tactile effect we observed was present only for the above-threshold tactile bias, and only as compared to the auditory pre- and post-tests, but not as compared to the no-bias tactile condition. Further because we have no evidence for a tactile effect in the "weak" bias condition, where the tactile stimulation was presented below threshold, we cannot exclude the possibility that subjects made use of the tactile information in a cognitive fashion, instead of integrating it perceptually with the auditory signal (cf. Massaro, 1987; Massaro and Cohen, 1983; Jain et al., 2010).

It might be argued that the reason we did not obtain interference from the below-threshold bias and only a small interference in the above-threshold bias was that subjects were not sufficiently trained: after all the association between our auditory and tactile stimuli was completely arbitrary, and it is known that in such cases extensive training is necessary to find an interaction.

However it should be noted that even if, with further training, we had obtained stronger effects, and in particular an effect of the sub-threshold tactile bias, we could still not be fully confident that the phenomenon was purely perceptual. Jain et al.'s (2010) argument about subliminal stimuli notwithstanding, even sub-threshold tactile information might at times during the experiment reach the subject's awareness and influence his or her decisions in a cognitive rather than perceptual way. For this reason we decided to use a tactile analogy of the McGurk effect (Experiment 2), which provides a perceptually obvious instance of multimodal integration. To avoid the problem of auditory adaptation due to repetition of the same syllables, we decided to increase the number of syllables we used.

## 3    Experiment 2: Audio-tactile McGurk

In the original McGurk effect, a visual /GA/ accompanied by an acoustic /BA/ is often perceived as /DA/; a visual /BA/ accompanied by an acoustic /GA/ is sometimes perceived as /BGA/ (McGurk & MacDonald, 1976). If we could find evidence for an audio-tactile McGurk type illusion, where a new syllable is perceived, then this would argue for true audio-tactile integration rather than unimodal selection of either the acoustic or the tactile stimulus (see 'bimodal speech perception' theory in auditory-visual domain, Massaro, 1987; Massaro and Cohen, 1983).

As far as we know, two previous studies have tried to demonstrate auditory-tactile integration using a paradigm close to the McGurk effect in the audio-tactile domain, pairing repeated auditory syllable to tactile stimulation (Fowler and Dekle, 1991; Sato et al. 2010). Both studies did not report a perceptual experience equivalent to the original McGurk effect (perception of a different syllable from the four presented). Fowler and Dekle (1991) found an equivalent of the McGurk effect in only one of 7 Subjects, even if there was evidence for a strong influence of the tactile stimulation on auditory perception of syllables (and vice versa). Sato et al. (2010) found no clear evidence for a McGurk type illusion, since they found perception of the same syllable in both unimodal and bimodal conditions. However, these two studies differ from ours for two reasons. First, they did not use an arbitrary code, as tactile aids do, but a method called Tadoma (Alcorn, 1932) where, in its original version, a deafblind person places one thumb lightly on the lips of the talker while the other fingers fan out over the face and neck. Secondly, in both studies there was no training phase. We hoped that, even though we used an arbitrary tactile stimulation code that Subjects had no prior familiarity with, after extensive training there might be evidence for auditory-tactile integration and that this integration would take place in a way similar to what happens in the original McGurk effect.

We therefore extensively trained two subjects to associate four syllables with four corresponding tactile patterns. In a subsequent test, we presented the subjects with the learnt auditory syllables, except now they were accompanied either with the previously associated tactile patterns (congruent trials), or with tactile patterns that were not associated with them (incongruent trials). We expected that in a way analogous to the McGurk effect, Subjects in the incongruent trials would perceive a completely "new" syllable or a syllable "inbetween" the auditory and tactile syllables.

### 3.1    Methods

#### 3.1.1    Subjects
2 male subjects (mean age 34.5 yrs), Russian and Italian native speakers (experimenters).

#### 3.1.2    Stimuli and Apparatus.
Auditory stimuli: We chose four syllables (/BA/, /VA/, /GA/, /DA/) that were easily confused, as measured by a classic confusion matrix for phonemes (Miller and Nicely, 1955). We used 10 different recordings of each syllable (40 stimuli overall). The syllables were presented at 4 different levels of volume ranging from just audible to very clearly audible : 25, 45, 55, 60 dB correspondingly.

Tactile stimuli: we used the same device as in Exp 1 and generated apparent motion in a similar way. Tactile stimulation was always perfectly easy to discriminate (time delay 60 ms, and 60 ms pin-up time, see Exp 1), so its direction could be easily judged by the subjects. We presented four distinguishable tactile stimuli using the external pins of the 4x4 array. The stimuli were: top row moving left, bottom row moving right, left column moving down, right column moving up (see Figure 6).

We trained the two subjects to associate each syllable to one tactile stroking direction. The subjects were offered two possible key mappings on the QWERTY keyboard for responding: one mapping corresponded to the position on the thumb that was being stimulated: W (for top row moving left and /DA/), A (for left column moving down and /BA/), S (for bottom row moving right and /GA/), D (for right column moving up and /VA/); the other mapping corresponded to the apparent motion of the pins: W (for right column moving up and /VA/), A (for top row moving left and /DA/), S (for left column moving down and /BA/), D (for bottom row moving right and /GA/). The association between auditory syllables and tactile stimuli was the same for both mappings. After a short trial subject 1 chose the former mapping and subject 2 chose the latter. In the test session both subjects could use the spacebar to respond "other" if they heard something different from one of the four stimuli. The Subjects responded with the left (non dominant) hand.

#### 3.1.3    Procedure
Each subject did daily training and a test after the last training session. For the training sessions (divided into, 'passive' 'exposure', and 'active' sessions), the subject sat in front of the HP Elitebook 14" computer wearing headphones with the thumb of his dominant (right) hand resting on the tactile device. Responses on the laptop keyboard were made with the non-dominant (left) hand. Subjects had a break after each session of 200 trials of training.

In the 'passive' and 'exposure' training sessions the subject was presented with one of the four auditory syllables accompanied by the tactile stimulation. In the 'passive' session the subject was instructed to respond using the allocated keys on the keyboard (see Stimuli), and the next stimulus was presented after the response. In the 'exposure' session the stimuli were presented continuously with 500 ms pauses. Subjects were instructed to mentally name the presented syllable. Both sessions included catch trials in which no tactile stimulation was presented. Each stimulation session was composed of 200 trials: 4 syllables x 10 instances x (4 amplitudes + 1 catch), the trials were randomized within session.

In the 'active' training session, subjects were instructed to verbally pronounce aloud the syllable presented visually on the computer screen. A custom computer program detected the onset of the voice with a latency under 50 ms, and generated the tactile stimulus that was associated with the visually presented syllable. The 'passive', 'exposure' and 'active' sessions were intermingled during the day.

The test session was composed of 36 blocks of ten trials each. All trials within the same block had the same auditory syllable and tactile pattern. For each syllable there were 9 blocks: 6 blocks with congruent tactile stimulation and 3 blocks with incongruent stimulation (1 block for each incongruent tactile pattern). For every syllable we had two instances of the audio recording. Each block contained five trials with each recording, their order was randomized within the block. During the session the blocks were presented in a random order. In total there were 360 trials: 240 congruent and 120 incongruent. We kept this imbalance in order to minimize the effect incongruent stimulations may have on the audio-tactile correspondence possibly learnt by the subjects.

The subjects were instructed to respond which syllable they have heard using the same keys as during the 'passive' session, and having an "other" option which was mapped into the space bar. For every block, the subjects were first exposed to five trials with 500 ms pause between the trials, which were followed by another five trials to which the subject responded using the keys. The subjects knew that the presentation came in blocks and that the trials within each block corresponded to the same audio and tactile stimuli, but were instructed to respond according to their perceptual experience. The motivation behind such presentation scheme was to give subjects time to focus on their perceptual experience. On total, 180 responses were collected.

Overall, subject 1 did 6 days of training, 4-5 hours per day (around 50,000 'passive' and 'exposure' trials and around 2,000 'active' trials), while subject 2 did 6 days of training, 1.5-2 hours per day (around 15,000 'passive' and 'exposure' trials and around 1,000 'active' trials). Each subject also ran one session of test per day (360 trials), usually immediately after the last training session, which always was 'passive'. In the test sessions only the two clearest recordings of each syllable were used instead of all ten recordings as used in the training sessions.

### 3.1.4 Results and Discussion

Our expectation was that for the incongruent trials subjects would perceive a 'new' syllable or a totally different syllable from the presented /BA/, /VA/, /DA/ or /GA/. However, nothing changed in subjects' auditory perception of the syllables either in the congruent or incongruent trials. In all trials the subjects correctly reported the auditory stimuli independently of the tactile stimulation. Interestingly, both subjects noticed that they became "immune" to the classical, visual McGurk effect soon after the study, meaning that they perceived veridical auditory syllable independently of the visual information. This immunity sustained for several months. No formal study has been performed and these observations should be considered anecdotal.

In this qualitative study we wanted to find proof of perceptual audio-tactile integration of auditory syllables with tactile cues. We hoped to find an equivalent of the McGurk effect, which we consider to be a good example of truly perceptual multisensorial integration. However even after the 6 day training we used, we found that tactile input from our device did not influence the auditory perception of syllables. We therefore have no evidence of truly perceptual audio-tactile integration.

## 4 Conclusion

In the introduction we pointed out that early tactile speech aids were initially partly successful, but that they never became viable and were abandoned. We suggested that one reason why these devices failed may have been that the tactile information provided by the prostheses was only integrated at a cognitive or decisional level, and could not be properly perceptually integrated into the auditory speech stream. This would have the consequence that too much cognitive effort would be required in order to use a tactile aid to understand speech delivered at normal speech rates, making such devices essentially unusable.

The purpose of the present experiments was therefore to see if we could show that it is actually possible to create a situation where a tactile cue is perceptually integrated with a speechlike auditory stimulus. To maximise our chance of success, Experiment 1 checked whether the perception of auditory syllables in an auditory continuum (from a clear /VA/ to a clear /BA/) would be influenced by simultaneous presence of a previously learnt compatible or incompatible tactile stimulation presented above and below threshold. We found that tactile stimulation modified the auditory discrimination of the syllables /BA/ and /VA/ only when the tactile stimulus was easily perceptible above threshold, but not when it was below threshold. Unfortunately, because facilitation only occurred when the tactile stimulus was consciously perceived, it is therefore possible that subjects used a strategy of selecting the most reliable unimodal consciously available information source, instead of experiencing true perceptual integration between the auditory and tactile modalities.

We therefore did a second, qualitative, study (Experiment 2) using a tactile analog of the McGurk effect to attempt to find proof of the existence of a truly perceptual effect of the tactile stimuli on the auditory percept. However, after six days' extensive training with a particular auditory-tactile association, we found that Subjects' auditory syllable perception did not change when the tactile stimulation was incongruent with the learned association. In other words, we did not find an audio-tactile equivalent of the audio-visual McGurk effect.

Neither of our experiments can thus be taken to provide evidence for the existence of truly perceptual auditory-tactile integration for the stimuli we used. Any integration we observed may possibly have been a kind of "selection strategy" (Fowel and Dekle, 1991), occurring at higher, "not properly perceptual" levels of processing.

It is always difficult to conclude from a negative result. Here it can obviously be claimed that perhaps with greater training or somehow different stimuli we might have been able to find the truly perceptual facilitation we were looking for. But the point of our study was to show that in conditions similar to those used in tactile aids, that is, with auditory quality simulating hearing impaired users who need to distinguish phonemes that their hearing makes ambiguous, and with approximately the amount of training that a patient would be willing to devote to the use of a tactile aid, integration is not easy to obtain. Even with six days intensive training attempting to link the tactile with the auditory stimuli, we failed to find any evidence of perceptual fusion in our second experiment.

As noted in the introduction, this finding is compatible with the historical fact that tactile devices never became commercially viable. It is also compatible with a careful reading of the literature on multimodal perception, where, contrary to first impressions, cognitive, decisional or response-level interpretations can often be invoked to explain any observed facilitation. Indeed in a series of studies summarized by Jain, Fuller & Backus (2010) and relating mostly to visual and auditory modalities, these authors and collaborators had undertaken to systematically explore under what conditions arbitrary, previously irrelevant cues can, through learning, be associated with a stimulus and bias its perceptual appearance. They concluded that some of the factors that may favour perceptual integration are: the fact that the cue is from the same modality as the stimulus; the fact that it is "intrinsic", i.e. part of the stimulus configuration itself; the fact that the cue has been associated with the stimulus over a life-long period; the fact that the cue is simultaneous with the stimulus; and possibly the fact that the cue does not contain too much information.

It is possible therefore that if the auditory stimulus we were seeking to facilitate had contained less information or had corresponded to a life-long learnt association, we would have succeeded. For example if we had simply wanted to facilitate the detection of the simple presence/absence of an auditory signal by the presence/absence of a tactile stimulation, then we might have succeeded.

However if we should not be surprised that if we wish to create a link between a set of several tactile codes and several corresponding phonemic features, as would be necessary in a tactile aid, then this will be much more difficult. Such codes would be arbitrary, cross-modal, correspond to no life-long familiar combinations, would not be "intrinsic", and therefore fail all the criteria suggested by Jain et al. (2010). Indeed, as suggested by Nava et al. (2014), it may even be the case that humans have a critical period in which such associations can be learnt, but outside this period, extremely long training would be necessary.

In conclusion, the failure of our experiments to demonstrate convincing perceptual integration provides a plausible explanation for the failure of the early tactile speech aids, and bodes ill for the future of such devices. This negative conclusion seems important to point out in the context of renewed efforts today to create sensory augmentation and brain machine interfaces[3].

---

[3] https://techcrunch.com/2017/04/19/facebook-brain-interface/ ; http://www.neosensory.com.

## 5 Acknowledgments

This work is funded by the ERC Grant FEEL Number 323674 and ERC Proof of Concept Grant FeelSpeech Number 692765. We wish to thank Ramakanth Singal Reddy for his important contribution in the construction of the device.

# 7    Figure legends

Fig.1 Tactile stimulation device

Fig. 2. Example of audio and tactile psychometric functions fitted to the data of two representative subjects (for the other subjects see Supplementary Material). For the audio graphs, each data point corresponds to 15 measurements. For the tactile graph (top, middle), each data point corresponds to 15 measurements. The means μ of each function are taken to be the PSE for that subject for that condition, and σ indicates the slope.

Fig. 3. Audio-tactile training. Auditory stimuli on the "/BA/ side" of the subject's auditory PSE were mapped into one particular tactile stimulus on the "moving left" side of the subject's tactile PSE. The tactile stimulus on the "moving left" side ranged from very clearly distinguishable (3σ from tactile PSE) at the beginning of the training to almost indistinguishable (0.5σ from tactile PSE) at the end of the training (only some of the intermediate mappings are shown in the Figure) . The "/VA/ side" auditory stimuli were mapped in a similar way into the "moving right" side tactile stimuli.

Fig. 4. The three conditions of the test. In the test block the mapping between auditory and tactile stimuli was identical to the last training session (tactile stimuli at PSE±0.5σ). In the biased conditions the mapping was shifted by 3 points towards /BA/. Weak and strong bias conditions differed by the strength of the tactile stimulation: ±0.5σ and ±2σ respectively.

Fig. 5. PSE values along the /VA/-/BA/ auditory continuum for each subject in each condition, and overall means. * = p<.05; ** = p<.005

Fig. 6. Stroking directions of the pins.

Supplementary Material

Fig. S1. Audio and tactile psychometric functions fitted to the data of the 10 subjects. For the audio graphs, each data point corresponds to 15 measurements. For the tactile graph (top, middle), each data point corresponds to 15 measurements. The means μ of each function are taken to be the PSE for that subject for that condition, and σ indicates the slope.

Audio files containing the stimuli numbers 0 to 20 with spliced combinations ranging from /VA/ to /BA/

Fig.1 Tactile stimulation device

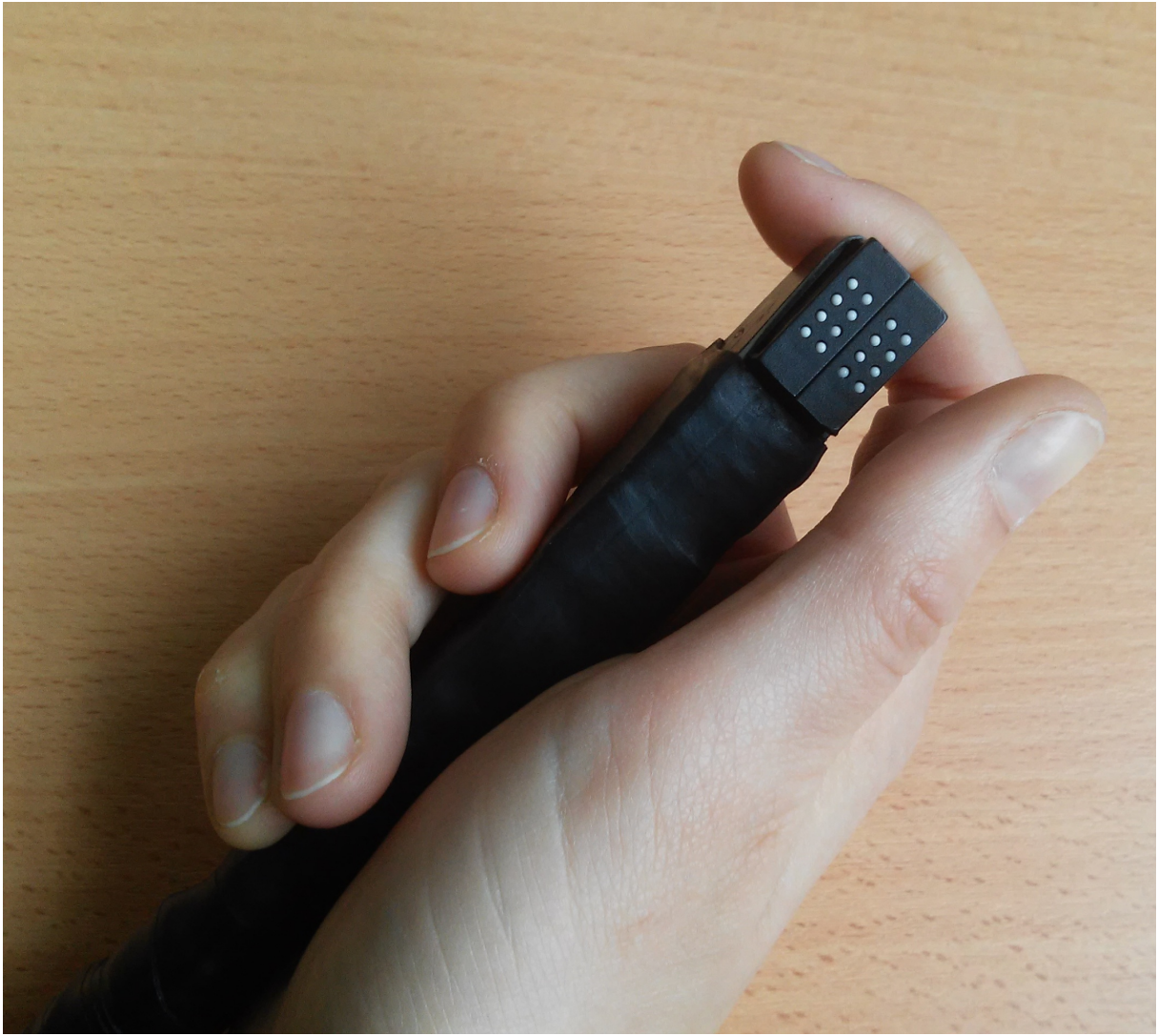

Fig. 2. Example of audio and tactile psychometric functions fitted to the data of two representative subjects (for the other subjects see Supplementary Material). For the audio graphs, each data point corresponds to 15 measurements. For the tactile graph (top, middle), each data point corresponds to 15 measurements. The means μ of each function are taken to be the PSE for that subject for that condition, and σ indicates the slope.

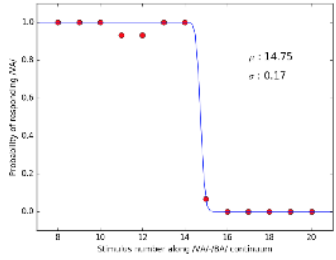
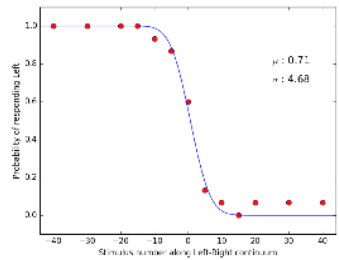
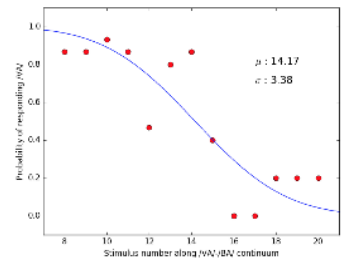
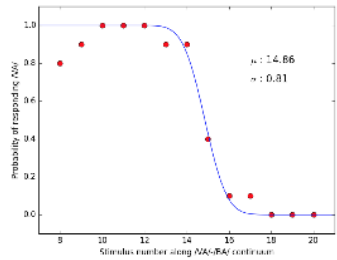
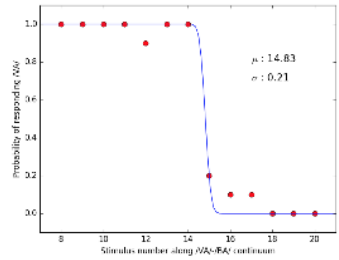
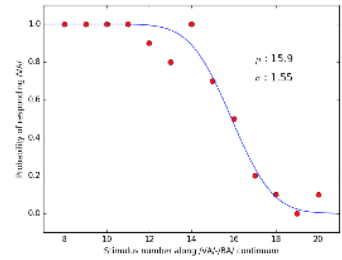
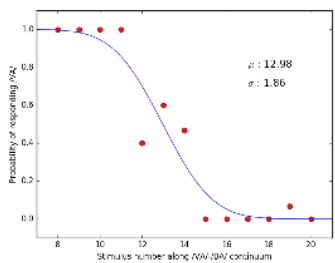
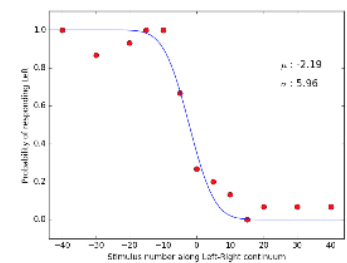
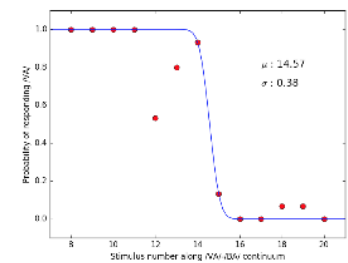
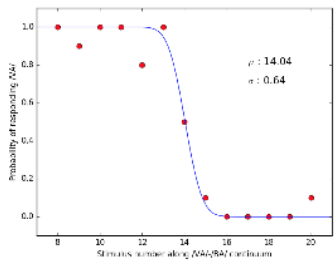
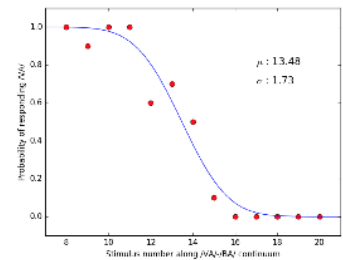
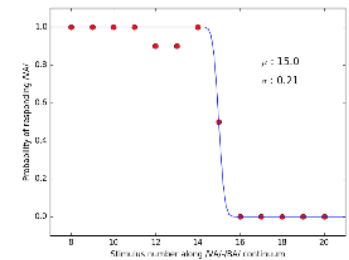

Fig. 3. Audio-tactile training. Auditory stimuli on the "/BA/ side" of the subject's auditory PSE were mapped into one particular tactile stimulus on the "moving left" side of the subject's tactile PSE. The tactile stimulus on the "moving left" side ranged from very clearly distinguishable (3σ from tactile PSE) at the beginning of the training to almost indistinguishable (0.5σ from tactile PSE) at the end of the training (only some of the intermediate mappings are shown in the Figure). The "/VA/ side" auditory stimuli were mapped in a similar way into the "moving right" side tactile stimuli.

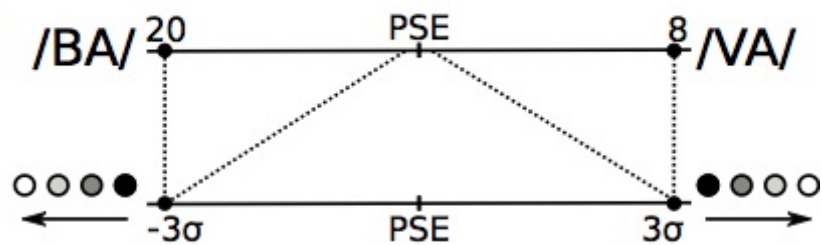

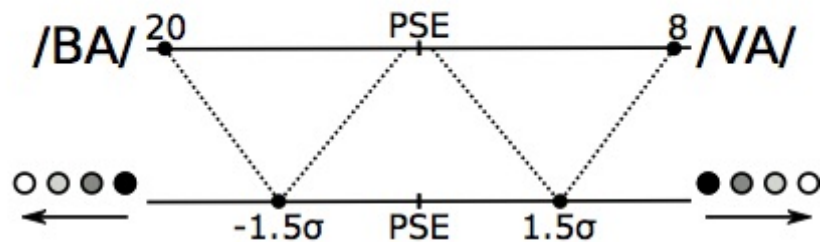

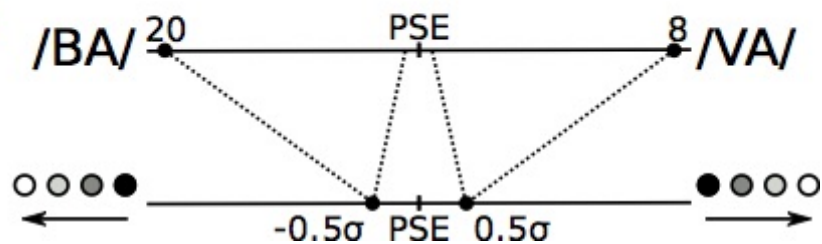

Fig. 4. The three conditions of the test. In the test block the mapping between auditory and tactile stimuli was identical to the last training session (tactile stimuli at PSE±0.5σ). In the biased conditions the mapping was shifted by 3 points towards /BA/. Weak and strong bias conditions differed by the strength of the tactile stimulation: ±0.5σ and ±2σ respectively.

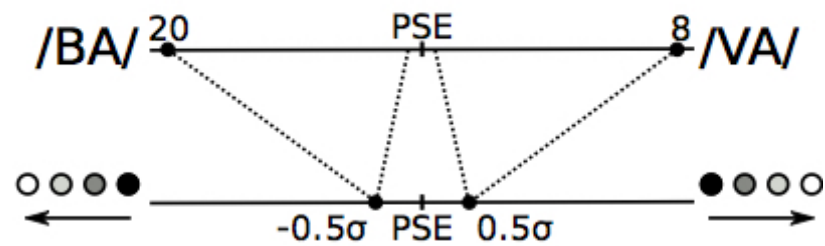
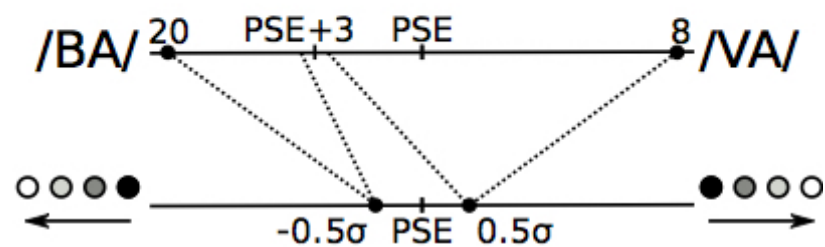
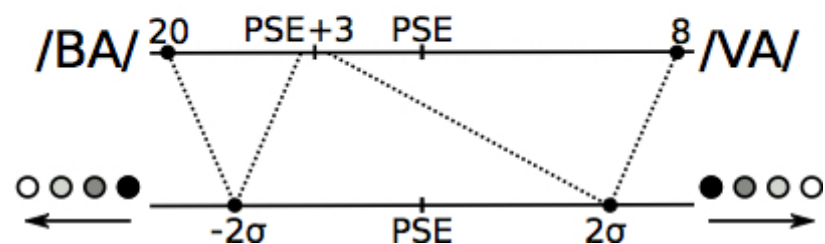

Fig. 5. PSE values along the /VA/-/BA/ auditory continuum for each subject in each condition, and overall means. * = p<.05; ** = p<.005

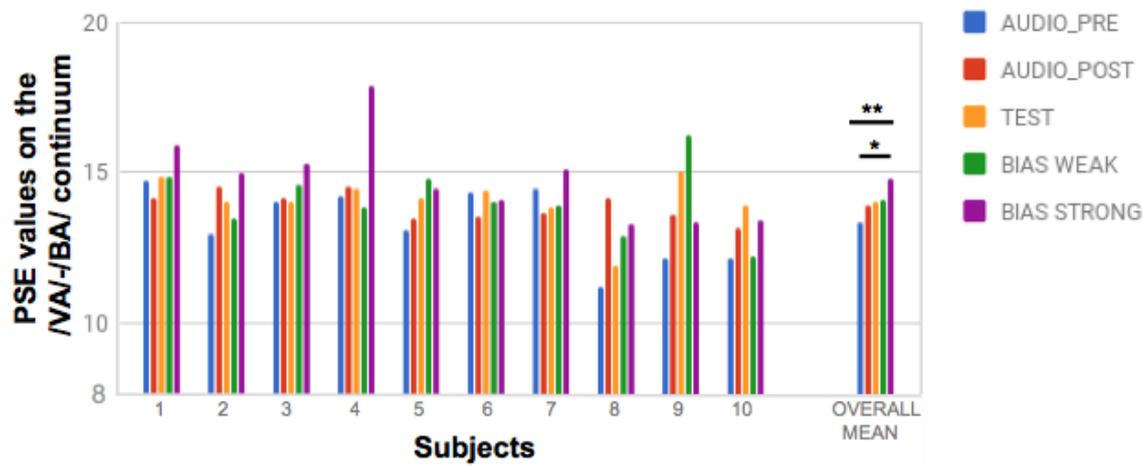

Fig. 6. Stroking directions of the pins.

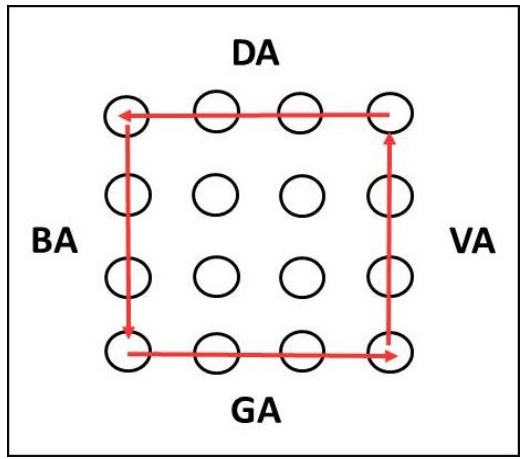

**Supplementary Material**

Audio files containing the stimuli numbers 0 to 20 with spliced combinations ranging from /VA/ to /BA/: cf https://github.com/montoneguglielmo/stimuliFeelSpeechFrontier

Fig. S1. Audio and tactile psychometric functions fitted to the data of the 10 subjects. For the audio graphs, each data point corresponds to 15 measurements. For the tactile graph (top, middle), each data point corresponds to 15 measurements. The means μ of each function are taken to be the PSE for that subject for that condition, and σ indicates the slope.

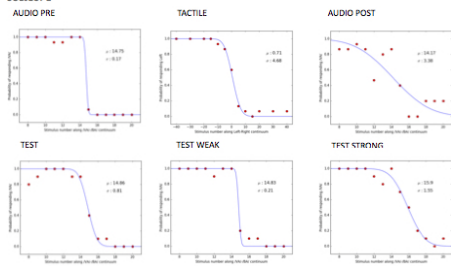
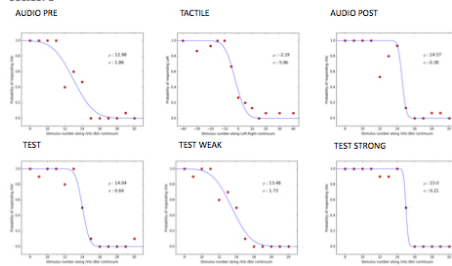
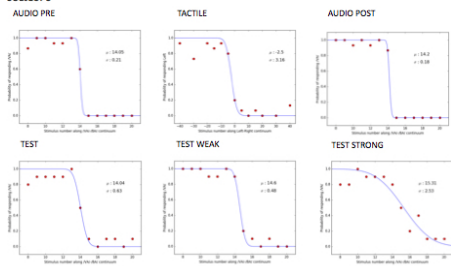
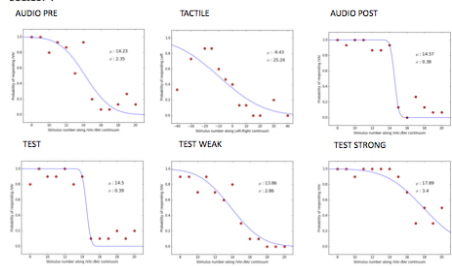
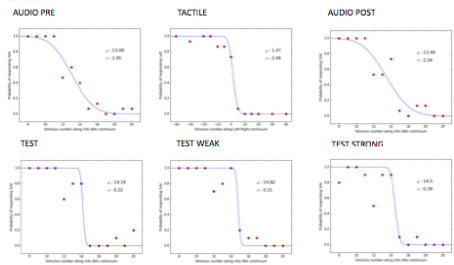
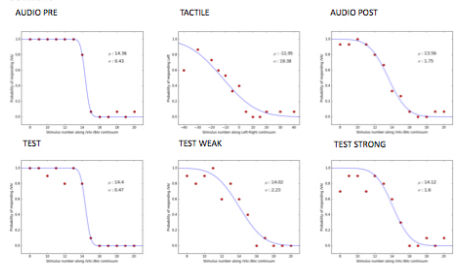

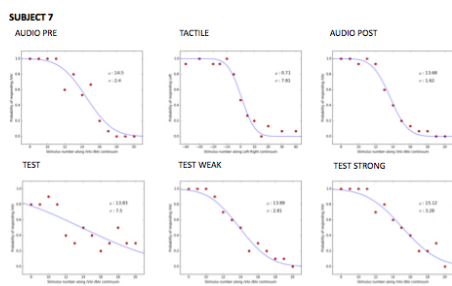
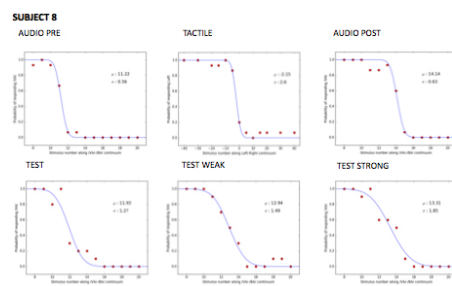
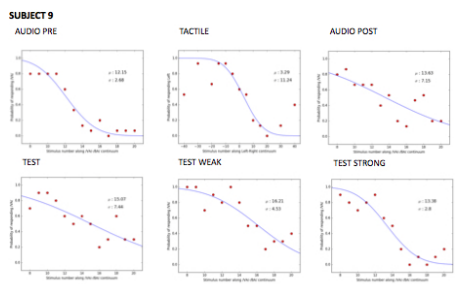
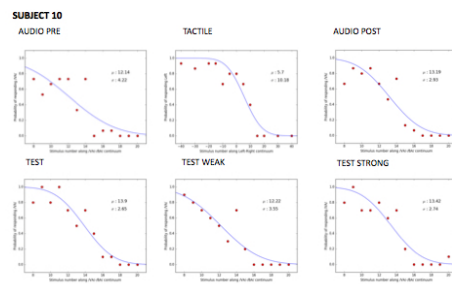